\documentclass{ptptex}

\newcommand{\be}{\begin{equation}}
\newcommand{\ee}{\end{equation}}
\newcommand{\bea}{\begin{eqnarray}}
\newcommand{\eea}{\end{eqnarray}}
\newcommand{\A}{{\mathbf A}}
\newcommand{\X}{{\mathbf X}}

\newcommand{\V}{{\mathbf V}}

\newcommand{\nn}{\nonumber}

\newcommand{\hn}{{\mathbf {\hat n}}}

\newcommand{\half}{\frac{1}{2}}

\title{Higgs-free confinement hierarchy in five colour QCD}
\author{Michael Luke \textsc{Walker}}
\inst{Institute of Quantum Science, College of Science and Technology,
Nihon University, Chiyoda, Japan 101-8308}

\abst{I consider the monopole condensate of five color QCD. 
The naive lowest energy state is unobtainable at one-loop for five
or more colors due to simple geometric considerations. The consequent
adjustment of the vacuum condensate generates a hierarchy of confinement scales in 
a natural Higgs-free manner. The accompanying symmetry hierarchy contains hints of
standard model phenomenology.}

\begin{document}

\maketitle

\subsection*{Introduction}
It is already known \cite{S77,F80,me07}, that $SU(N)$ QCD 
can lower the energy of its vacuum with
a monopole background field along the Abelian directions, where the Abelian components are
equal in magnitude but orthogonal in real space \cite{F80,me07}. This orthogonality,
while of no special consequence in $SU(3)$ QCD in three space 
dimensions, does have consequences
when the number of Abelian directions is greater than three.
As noted originally by Flyvberg \cite{F80}, $SU(N \ge 5)$ QCD cannot realise its true 
minimum because four orthogonal vectors cannot fit in three
dimensions. I shall call a system kept from reaching its true lowest
energy state by a lack of spatial dimensions \emph{dimensionally frustrated}. 

The effective ground state energy of the monopole condensate is a non-analytic
expression in the Cartan components $\mathbf{H}^{(i)}$. However, its form 
strongly suggests that the lowest physically available energy state is the
one that leaves the effective condensate felt by each of the off-diagonal (valence)
gluons equal in magnitude, or as close to it as can be realised. This is the 
motivation for my condensate ansatz. The lowest
energy state allowed by it leaves the (matter-antimatter pair of) gluons associated
with one root vector
very strongly confined, another very weakly confined, and the rest confined with 
identical
intermediate strength. Examining the dynamics at intermediate energy scales where 
the strongly confined dynamics have dropped out finds an emergent three-colour QCD
accompanied by gluons whose dynamics no longer conform to a special unitary symmetry
group. They do however form two six-dimensional representations of $SU(3)$. One of these
representations consists of the weakly confined valence gluons. Exactly one unconfined 
massless Abelian gauge field (photon) can be found by taking
linear combinations of the accompanying Abelian gluons.

\subsection*{The Cho-Faddeev-Niemi decomposition}
My treatment of the monopole condensate rests on tbe Cho-Faddeev-Niemi (CFN)
decomposition \cite{Cho80a,FN99c,LZZ00}. I use the following notation: \newline
The Lie group $SU(N)$ has $N^2-1$ generators $\lambda^{(j)}$, of which $N-1$
are Abelian generators $\Lambda^{(i)}$. 
For simplicity, we specify the 
gauge transformed Abelian directions (Cartan generators)
with 
\be
\hn_i = U^\dagger \Lambda^{(i)} U. 
\ee
In the same way, we replace the standard
raising and lowering operators $E_{\pm\alpha}$ for the root vectors $\mathbf{\alpha}$ 
with the gauge transformed ones
\be
E_{\pm \alpha} \rightarrow U^\dagger E_{\pm \alpha} U,
\ee
where $E_{\pm \alpha}$ refers to the gauge transformed operator 
throughout the rest of this article.

Gluon fluctuations in the $\hn_i$ directions are described by $c^{(i)}_\mu$. 
The gauge field of the covariant derivative which leaves
the $\hn_i$ invariant is
\bea
g\mathbf{V}_\mu \times \hn_i = -\partial_\mu \hn_i.
\eea
In general this is 
\bea
\mathbf{V}_\mu = c^{(i)}_\mu \hn_i + \mathbf{B}_\mu ,\; \;
\mathbf{B}_\mu = g^{-1} \partial_\mu \hn_i \times \hn_i,
\eea
where summation is implied over $i$. $\mathbf{B}_\mu$ can be a attributed to
non-Abelian monopoles, as indicated by the $\hn_i$ describing the homotopy group
$\pi_2[SU(N)/U(1)^{\otimes (N-1)}] = \pi_1[U(1)^{\otimes (N-1)}]$.
The monopole field strength
\be
\mathbf{H}_{\mu \nu} = \partial_\mu \mathbf{B}_\nu - \partial_\nu \mathbf{B}_\mu
+ g\mathbf{B}_\mu \times \mathbf{B}_\nu,
\ee
has only Abelian components, \textit{ie}. 
\be
H^{(i)}_{\mu\nu}\,\hn_i = \mathbf{H}_{\mu\nu},
\ee
where $H^{(i)}_{\mu\nu}$ has the eigenvalue $H^{(i)}$. Since I am only
concerned with magnetic backgrounds, $H^{(i)}$ is considered the magnitude
of a background magnetic field $\mathbf{H}^{(i)}$. The field strength of the Abelian
components $c_\mu^{(i)}$ also lies in the Abelian directions as expected and is shown
by
\bea
\mathbf{F}_{\mu\nu} = F_{\mu\nu}^{(i)} \hn_i, 
\eea
Defining
\be
F^{(i)}_{\mu\nu} = \partial_\mu c_\nu^{(i)} - \partial_\nu c_\mu^{(i)},
\ee
the Lagrangian of the Abelian and monopole components is
\bea \label{eq:Abelian}
-\frac{1}{4} (F_{\mu\nu}^{(i)} \hn_i + \mathbf{H}_{\mu\nu})^2
\eea

The dynamical degrees of freedom (DOF) perpendicular to $\hn_i$ are denoted by
$\X_\mu$, so if $\A_\mu$ is the gluon field then
\bea
\A_\mu &=& \V_\mu + \X_\mu 
= c^{(i)}_\mu \hn_i + \mathbf{B}_\mu + \X_\mu,
\eea
where
\bea
\X_\mu \bot \hn_i , \; \;
\X_\mu = g^{-1}\hn_i \times \mathbf{D}_\mu \hn_i, \; \;
\mathbf{D}_\mu = \partial_\mu + g\A_\mu \times. 
\eea
Because $\X_\mu$ is orthogonal to all Abelian directions it can be expressed as 
a linear combination of the raising and lowering operators $E_{\pm\alpha}$, which
leads to the definition
\begin{eqnarray}
X_\mu^{(\pm \alpha)} \equiv E_{\pm\alpha} \mbox{Tr}[\mathbf{X}_\mu E_{\pm\alpha}],
\end{eqnarray}
so
\begin{equation}
X_\mu^{(-\alpha)} = {X_\mu^{(+ \alpha)}}^\dagger .
\end{equation}

$\mathbf{H}_{\mu \nu}^{(\alpha)}$, defined by
\be
\mathbf{H}_{\mu \nu}^{(\alpha)} = \alpha_j H_{\mu \nu}^{(j)},
\ee
is the monopole field strength tensor felt by $\mathbf{X}_\mu^{(\alpha)}$.
I also define the background magnetic field
\be
\mathbf{H}^{(\alpha)} = \alpha_j \mathbf{H}^{(j)},
\ee
whose magnitude $H^{(\alpha)}$ is $\mathbf{H}_{\mu \nu}^{(\alpha)}$'s non-zero eigenvalue.

\subsection*{The Vacuum State of five-color QCD}

The one-loop effective energy of five-color QCD is given by \cite{F80,me07}
\bea
\label{eq:su4ground}
\mathcal{H} = \sum_{\alpha>0} \Vert \mathbf{H}^{(\alpha)} \Vert^2 
\left[\frac{1}{5g^2} + \frac{11}{48\pi^2}
\ln \frac{H^{(\alpha)}}{\mu^2} \right]
\eea
which is minimal when
\bea
H^{(\alpha)} = \mu^2 \exp \left(-\half - \frac{48\pi^2}{55g^2} \right).
\eea
This neglects an alleged imaginary component \cite{NO78} which 
has been called into serious question recently 
\cite{H72,CmeP04,Cme04,CP02,K04,KKP05,me07} with growing evidence to suggest that
it is only an artifact of the quadratic approximation. Taking this to be the case,
I employ the Savvidy vacuum. This can be criticised for lacking Lorentz covariance
but I argue that it is likely to match the true vacuum at least locally.

Since
\bea \label{eq:rootmagnitudes}
\Vert \mathbf{H}^{(1,0,0,0)} \Vert &=& \Vert \mathbf{H}^{(1)} \Vert ,\nn \\
\Big\Vert \mathbf{H}^{\left(\pm\frac{1}{2},\frac{\sqrt{3}}{2},0,0\right)} \Big\Vert^2 
&=& \frac{1}{4} \Vert \mathbf{H}^{(1)} \Vert^2 + \frac{3}{4} \Vert \mathbf{H}^{(2)} \Vert^2
\pm \frac{\sqrt{3}}{2} \mathbf{H}^{(1)} \cdot \mathbf{H}^{(2)} , \nn \\
\Big\Vert \mathbf{H}^{\left(\pm\frac{1}{2},\frac{1}{\sqrt{12}},\frac{2}{\sqrt{6}},0\right)} \Big\Vert^2 
&=& \frac{1}{4} \Vert \mathbf{H}^{(1)} \Vert^2 + \frac{1}{12} \Vert \mathbf{H}^{(2)} \Vert^2
+ \frac{2}{3} \Vert \mathbf{H}^{(3)} \Vert^2 
\pm \sqrt{\frac{2}{3}} \mathbf{H}^{(1)} \cdot \mathbf{H}^{(2)} \nn \\
&&\pm \frac{1}{2\sqrt{3}} \mathbf{H}^{(1)} \cdot \mathbf{H}^{(3)}
+ \frac{\sqrt{2}}{3} \mathbf{H}^{(2)} \cdot \mathbf{H}^{(3)}, \nn \\
\Big\Vert \mathbf{H}^{\left(0,-\frac{1}{\sqrt{3}},\frac{2}{\sqrt{6}},0\right)} \Big\Vert^2 
&=& \frac{1}{3} \Vert \mathbf{H}^{(2)} \Vert^2 + \frac{2}{3} \Vert \mathbf{H}^{(3)} \Vert^2
- \frac{2\sqrt{2}}{3} \mathbf{H}^{(2)} \cdot \mathbf{H}^{(3)}, \nn \\
\Big\Vert \mathbf{H}^{\left(0,0,-\frac{\sqrt{3}}{\sqrt{8}},\frac{\sqrt{5}}{\sqrt{8}}\right)} \Big\Vert^2 
&=& \frac{3}{8} \Vert \mathbf{H}^{(3)} \Vert^2 + \frac{5}{8} \Vert \mathbf{H}^{(4)} \Vert^2
- \frac{\sqrt{15}}{4} \mathbf{H}^{(3)} \cdot \mathbf{H}^{(4)} , \nn \\
\Big\Vert \mathbf{H}^{\left(0,-\frac{\sqrt{3}}{\sqrt{8}},\frac{1}{\sqrt{24}},\frac{\sqrt{5}}{\sqrt{8}}\right)} \Big\Vert^2 
&=& \frac{3}{8} \Vert \mathbf{H}^{(2)} \Vert^2 + \frac{1}{24} \Vert \mathbf{H}^{(3)} \Vert^2
+ \frac{\sqrt{5}}{\sqrt{8}} \Vert \mathbf{H}^{(4)} \Vert^2 
- \sqrt{\frac{1}{16}} \mathbf{H}^{(2)} \cdot \mathbf{H}^{(3)} \nn \\
&&- \frac{\sqrt{15}}{4} \mathbf{H}^{(2)} \cdot \mathbf{H}^{(4)}
+ \frac{\sqrt{5}}{\sqrt{48}} \mathbf{H}^{(3)} \cdot \mathbf{H}^{(4)}, \nn \\
\Big\Vert \mathbf{H}^{\left(\pm\frac{1}{2},\frac{1}{\sqrt{12}},\frac{1}{\sqrt{24}},
\frac{\sqrt{5}}{\sqrt{8}}\right)} \Big\Vert^2 
&=& \frac{1}{4} \Vert \mathbf{H}^{(1)} \Vert^2 + \frac{1}{12} \Vert \mathbf{H}^{(2)} \Vert^2
+ \frac{1}{24} \Vert \mathbf{H}^{(3)} \Vert^2 + \frac{\sqrt{5}}{\sqrt{8}} \Vert \mathbf{H}^{(4)} \Vert^2 \nn \\
&&\pm \sqrt{\frac{2}{3}} \mathbf{H}^{(1)} \cdot \mathbf{H}^{(2)} 
\pm \frac{1}{\sqrt{24}} \mathbf{H}^{(1)} \cdot \mathbf{H}^{(3)}
+ \frac{1}{\sqrt{16}} \mathbf{H}^{(2)} \cdot \mathbf{H}^{(3)} \nn \\
&&\pm \frac{\sqrt{5}}{\sqrt{8}} \mathbf{H}^{(1)} \cdot \mathbf{H}^{(4)}
+ \frac{\sqrt{5}}{\sqrt{24}} \mathbf{H}^{(2)} \cdot \mathbf{H}^{(4)}
+ \frac{\sqrt{5}}{\sqrt{48}} \mathbf{H}^{(3)} \cdot \mathbf{H}^{(4)},
\eea
it follows that 
\be
\Vert \mathbf{H}^{(i)}\Vert = \Vert \mathbf{H}^{(j)}\Vert ,\;\;
\mathbf{H}^{(i)} \bot \mathbf{H}^{(j)} ,\;\; i\ne j,
\ee
which means that
the chromomagnetic field components must be equal in magnitude but
mutually orthogonal in the lowest energy state. However three dimensional space can
only accomodate three mutually orthogonal vectors. Since the number of Cartan components
is always $N-1$ in $SU(N)$ it follows that QCD with more than four colors cannot 
achieve such an arrangement.

One could substitute the Cartan basis $\mathbf{H}^{(i)}$ but this leads
to intractable equations that cannot be solved analytically. It is reasonable to expect that
the lowest attainable energy state is only slightly different from (\ref{eq:su4ground}) and
that this difference is due to the failure of mutual orthogonality. I therefore propose the
ansatz that all Cartan components are equal in magnitude to what they would be in the absence
of dimensional frustration, and that their relative orientations in real space are chosen so as
to minimise the energy. In practice this means that three of the four are mutually orthogonal
and the remaining one is a linear combination of those three. This remainder will increase the 
effective energy through its scalar products with the mutually orthogonal vectors but not all
scalar products contribute equally. This follows from the form of the root vectors in 
eq.~(\ref{eq:rootmagnitudes}). This means that the orientation of the remaining real space
vector in relation to the mutually orthogonal ones impacts the effective energy. 

A little 
thought
reveals that the lowest energy state should have only one scalar product contribute to it. The
problem of finding the lowest available energy state therefore reduces to finding the scalar product 
that contributes to it the least. The six candidates are
\bea \label{eq:candidates}
\mathbf{H}^{(1)} \cdot \mathbf{H}^{(2)} ,\mathbf{H}^{(1)} \cdot \mathbf{H}^{(3)} ,\mathbf{H}^{(1)} \cdot \mathbf{H}^{(4)} ,\nn \\
\mathbf{H}^{(2)} \cdot \mathbf{H}^{(3)} ,\mathbf{H}^{(2)} \cdot \mathbf{H}^{(4)} ,\mathbf{H}^{(3)} \cdot \mathbf{H}^{(4)} .
\eea
\begin{table}
	\centering
	\caption{Candidate parallel components for vacuum condensate. The column on the left 
is for parallel vectors, the column on the right is for antiparallel vectors.
$\Delta \mathcal{H}$ should be multiplied by $H^2 \frac{11}{96\pi^2}$.}
		\begin{tabular}{cc|cc} \hline \hline
			$\mathbf{H}^{(i)} = + \mathbf{H}^{(j)}$ & $\Delta \mathcal{H}$ 
			& $\mathbf{H}^{(i)} = - \mathbf{H}^{(j)}$ & $\Delta \mathcal{H}$ \\ \hline
			$\mathbf{H}^{(1)} = + \mathbf{H}^{(2)}$ & $1.06381$ &
			$\mathbf{H}^{(1)} = - \mathbf{H}^{(2)}$ & $1.06381$ \\
			$\mathbf{H}^{(1)} = + \mathbf{H}^{(3)}$ & $0.857072$ &
			$\mathbf{H}^{(1)} = - \mathbf{H}^{(3)}$ & $0.857072$ \\
			$\mathbf{H}^{(1)} = + \mathbf{H}^{(4)}$ & $0.715651$ &
			$\mathbf{H}^{(1)} = - \mathbf{H}^{(4)}$ & $0.715651$ \\
			$\mathbf{H}^{(2)} = + \mathbf{H}^{(3)}$ & $1.01655$ &
			$\mathbf{H}^{(2)} = - \mathbf{H}^{(3)}$ & $0.656584$ \\
			$\mathbf{H}^{(2)} = + \mathbf{H}^{(4)}$ & $0.882589$ &
			$\mathbf{H}^{(2)} = - \mathbf{H}^{(4)}$ & $0.577976$ \\
			$\mathbf{H}^{(3)} = + \mathbf{H}^{(4)}$ & $1.00042$ &
			$\mathbf{H}^{(3)} = - \mathbf{H}^{(4)}$ & $0.540983$ \\ \hline
		\end{tabular}
	\label{tab:Candidates}
\end{table}
As can be seen from table \ref{tab:Candidates},
$\mathbf{H}^{(3)} = -\mathbf{H}^{(4)}$ (antiparallel) 
yields the lowest effective energy when all other scalar products are zero. 

Substituting this result into (\ref{eq:rootmagnitudes}) finds that all $\mathbf{H}^{(\alpha)}$ 
have the same magnitude except for those that couple to $\mathbf{H}^{(4)}$, namely
$\mathbf{H}^{\left(?,?,?,\sqrt{\frac{5}{8}}\right)}$, 
where ? indicates that there are several possible values.
The other background field strengths are 
\be
\Vert \mathbf{H}^{(\alpha)} \Vert^2 = H^2,
\ee
while the strongest is
\be
\Vert \mathbf{H}^{\left(0,0,-\sqrt{\frac{3}{8}},\sqrt{\frac{5}{8}}\right)} \Vert^2 
= H^2 \left(1+\frac{\sqrt{15}}{4}\right),
\ee
and the weakest are
\be
\Vert \mathbf{H}^{\left(?,?,\frac{1}{\sqrt{24}},\sqrt{\frac{5}{8}}\right)} \Vert^2 
= H^2 \left(1-\sqrt{\frac{5}{48}} \right).
\ee
Remember the negative signs are affected by $\mathbf{H}^{(3)},\mathbf{H}^{(4)}$ being 
antiparallel.

Assuming the dual superconductor model of confinement \cite{N74,M76,P77,Cho80a,tH81},
it follows that different valence gluons and even different quarks (in the fundamental representation)
will be confined with different strengths and therefore at different length scales. Those that feel the 
background $H^{\left(0,0,-\frac{\sqrt{3}}{\sqrt{8}},\frac{\sqrt{5}}{\sqrt{8}}\,\right)}$ will 
be confined the most strongly, those that feel the backgrounds of the form
$H^{\left(?,?,\frac{1}{\sqrt{24}},\frac{\sqrt{5}}{\sqrt{8}}\,\right)}$ will be confined
least strongly, where ? indicates that there are several values. The remainder will be confined with intermediate strength.
The confining potential felt by quarks is fundamentally different from that of the valence gluons \cite{KT00a,KT00b,me07}
and will not be treated in this letter.

At highest energy then, we have the full dynamics of $SU(5)$ QCD. Moving down to some intermediate energy
however, finds that the dynamics associated with the root vector
$\left(0,0,-\frac{\sqrt{3}}{\sqrt{8}},\frac{\sqrt{5}}{\sqrt{8}}\right)$
are confined out of the dynamics. The remaining
gluons interact among themselves. Moving to lower energy scales we find that those
dynamics are all removed in their turn except for those corresponding to the root vectors
$\left(?,?,\frac{1}{\sqrt{24}},
\frac{\sqrt{5}}{\sqrt{8}}\right)$, almost leaving an $SU(2)$ gauge field interaction.
I say 'almost' because I shall later demonstrate that the form of the monopole condensate
is sufficiently different from the $SU(2)$ condensate to alter the dynamics, producing three
confined $U(1)$ gauge fields, one unconfined
$U(1)$ gauge field that may be identified with the photon, and three copies of 
the valence gluons of $SU(2)$. 
At lowest energies only the unconfined gauge field remains.
In this way a hierarchy of confinement scales and effective dynamics emerges
naturally, without the introduction of any \textit{ad.~hoc.} mechanisms like the
Higgs field.
\subsection*{Intermediate Energy Dynamics}
In constructing the heirachical picture above, we began with $SU(5)$ and finished
with $U(1)$ but had no apparent gauge group governing the dynamics
at the intermediate energy scale. The dynamics of this energy scale will prove to be 
quite interesting. 

To facilitate the discussion I introduce a notation inspired by the Dynkin diagram
of $SU(5)$. The root vectors implicitly specified in eq.~(\ref{eq:rootmagnitudes}) are
all linear combinations of a few basis vectors, which according to Lie algebra representation
theory can be chosen for convenience. I take the basis vectors 
\bea
(1,0,0,0),\left(-\half,\frac{\sqrt{3}}{2},0,0 \right),
\left(0,-\frac{1}{\sqrt{3}},\sqrt{\frac{2}{3}},0 \right),
\left(0,0,-\sqrt{\frac{3}{8}},\sqrt{\frac{5}{8}}\right),
\eea
which I shall each represent by 
\be
\mbox{OXXX},\mbox{XOXX},\mbox{XXOX},\mbox{XXXO},
\ee
respectively. The remaining root vectors are sums of these basis vectors. In this notation
their representation contains an 'O' if the corresponding basis vector is included and
'X' if it is not. For example the root vector
\be
\left(\half,\frac{\sqrt{3}}{2},0,0 \right) 
= (1,0,0,0) + \left(-\half,\frac{\sqrt{3}}{2},0,0\right),
\ee
is represented by
\be
\mbox{OOXX}=\mbox{OXXX}+\mbox{XOXX}.
\ee
When convenient, a '?' is used to indicate that either 'O' or 'X' might be substituted.

In addition to its brevity, this notation has the nice feature of making obvious which
root vectors can be combined to form other root vectors because there are
no root vectors with an 'X' with 'O's on either side. There is no OXXO for example.

The confinement of $X_\mu^{\left(0,0,-\frac{\sqrt{3}}{\sqrt{8}},
\frac{\sqrt{5}}{\sqrt{8}}\right)}$, the valence gluon corresponding to XXXO, 
out of the dynamics 
directly affects only those remaining valence gluons that couple to it, those of root vectors of
the form ??O?. The remaining gluons, corresponding to 
the root vectors OOXX, XOXX and OXXX (collectively given by ??XX), 
may still undergo the full set of
interactions available to them at higher energies. It is easy to see that these are the
root vectors that comprise the group $SU(3)$, to which the other valence gluons couple
forming two six dimensional representations. 
Subsequent discussion shall
extend the X,O,? notation to include the valence gluons corresponding to a root vector.
Whether it is the gluon or the root vector that is meant will be clear from context.

Consider the beta function, or to be less imprecise, the 
scaling of the various gluon couplings. I shall now demonstrate that the loss to 
confinement of the root vector XXXO causes unequal corrections to the running of the
couplings for different gluons. 
Since this is only an introductory paper the following analysis is
only performed to one-loop.

The gluons ??XX, corresponding to the above-mentioned $SU(3)$, retain their original 
set of interactions. Performing the standard perturbative calculation \cite{F80}
therefore yields the standard result for $SU(5)$ QCD. The remaining gluons do not.
The absence of the maximally confined XXXO restricts their three-point vertices to those of
$SU(4)$, since all root vectors are now of the form ???X.
The same is not true of the four-point interactions, but the exceptions do
not contribute to the scaling of the coupling constant at one-loop \cite{Fbook87}. 
We have then that the
$SU(3)$ subgroup's coupling scales differently from the rest of the unconfined gluons when the
maximally confined valence gluons XXXO drop out.

The beta function is proportional to the number of colors in pure QCD at one-loop,
so as the length scale increases, the coupling among gluons within the $SU(3)$ subgroup
initially grows faster than the couplings involving the other gluons. As noted above,
the $SU(3)$ couplings will initially scale as in the five-color theory, 
while the remainder scale as though there were only four colors.
This specific behaviour must soon change due to both non-perturbative contributions
and because the non-$SU(3)$ gluons have a weaker coupling. A detailed understanding
requires a nonperturbative analysis well beyond the scope of this letter. 
Indeed, the application of one-loop perturbation 
theory at anything other than the far ultraviolet is questionable in itself. The point
remains that the $SU(3)$ subgroup X??X separates from the remaining 
gluons by its stronger coupling strength.

The symmetry reduction that takes place in this model is 
suggestive of boson mass generation but there appears to be no obvious
specific mechanism. Kondo \textit{et.~al.} have argued for the spontaneous generation
of mass through various non-trivial mechanisms \cite{K04,K06,me07}. This is consistent
with the well-studied correlation between confinement and chiral symmetry breaking
(see \cite{HF04,KLP01,YS82,PW84} and references therein).

\subsection*{The emergence of QED}
Neglecting off-diagonal gluons, 
the equality $\mathbf{H}^{(3)}=-\mathbf{H}^{(4)}$ allows the change in variables
\bea \label{eq:AandZ} 
c_\mu^{(3)}\hn_3 \rightarrow \half (c_\mu^{(3)}\hn_3 + c_\mu^{(4)}\hn_4) 
+ \half (c_\mu^{(3)}\hn_3 - c_\mu^{(4)}\hn_4) 
= \frac{1}{\sqrt{2}}(A_\mu + Z^0_\mu) ,\nn \\
c_\mu^{(4)}\hn_4 \rightarrow \half (c_\mu^{(3)}\hn_3 + c_\mu^{(4)}\hn_4) 
- \half (c_\mu^{(3)}\hn_3 - c_\mu^{(4)}\hn_4)
= \frac{1}{\sqrt{2}}(A_\mu - Z^0_\mu ).
\eea
Substituting eqs~(\ref{eq:AandZ}) into the Abelian dynamics (\ref{eq:Abelian})
finds that the antisymmetric combination $Z^0_\mu$
couples to the background 
\begin{displaymath}
H (\hn_3 - \hn_4),
\end{displaymath}
but the symmetric combination $A_\mu$
does not. Again by the dual superconductor model the former is confined (along with
$c_\mu^{(1)}\hn_1,c_\mu^{(2)}\hn_2$) while the latter
is not. Since neutral weak currents are short range and the electromagnetic field is
long range the natural interpretation of these combinations are the $Z^0$ for the 
symmetric combination and the photon for the antisymmetric combination.

The rotation from $c_\mu^{(3)}\hn_3,c_\mu^{(4)}\hn_4$ to 
$Z^0_\mu,A_\mu$ in interactions with
valence gluons is only meaningful if the gluon in question couples to both
$c_\mu^{(3)}\hn_3$ and $c_\mu^{(4)}\hn_4$. Otherwise the combination of $Z^0_\mu$ and 
$A_\mu$ is ill-defined because it is not unique, \textit{ie.} if the valence gluon  
couples to $c_\mu ^{(3)}\hn_3$ but not to $c_\mu^{(4)} \hn_4$ then arbitrary multiples of
$c_\mu^{(4)} \hn_4$ may be added to the interaction term, yielding
arbitrary mixtures of $Z^0_\mu$ and $A_\mu$. The 
gluon for which this occurs are of the form ??OX. Consequently the concept of coupling to
$A_\mu$ with a conserved charge is only meaningful in the low energy effective theory 
in which all ??OX have been confined out of the dynamics.

\subsection*{Summary} I have studied the long known but generally ignored result that
QCD with five or more colors has an altered vacuum state due to the limited 
dimensionality of space, a condition dubbed 'dimensional frustration'. Attempting
to identify the physical vacuum encounters an intractable set of non-analytic 
equations but a well-motivated ansatz enabled further analysis. Assuming the dual
superconductor model, a range of confinement
scales emerged with one root vector being confined more strongly than all the rest,
and others less tightly. The remaining gluons exhibit unconventional dynamics
at intermediate energy scales
because only some of them couple to the XXXO, which is most strongly confined.
At intermediate energies, a subset of these intermediate gluons represent $SU(3)$ 
and have stronger interactions among themselves.

The intermediate $SU(3)$ symmetry, the low energy $SU(2)$s, and the single unconfined
photon are tantalising hints of standard model phenomenology,
but this work is a long way from having reproduced it. 
If future work along these lines does reproduce it, the $W^\pm_\mu$
would be identified with the off-diagonal generators of ??OO. These interact with the
$SU(3)$ subgroup ??XX, corresponding to direct interaction between the $W^\pm_\mu$
and the QCD gluons, which does not occur in the standard model. However to my knowledge,
it has never been experimentally tested either. It predicts anomolous scattering of $W^\pm_\mu$
when fired at deep inelastic scattering energies into proton targets.

It could be
reasonably objected that the $W^\pm_\mu$ and $Z^0_\mu$ are 
not confined, but this is simply the current
understanding of the standard model. Experimentally we know that they have only short
lifetimes and never observed to propagate freely over significant distances. As such it
can be argued that they are confined, but very loosely.

There is considerable work to follow from the humble beginning presented here.
The dynamics of the fundamental representation have yet to be studied. Recent work on
the non-Abelian Stokes' theorem \cite{KT00a,KT00b} demonstrates that quark confinement is not 
synonomous with gluon confinement. It would be particularly interesting
to see whether a non-confined pair of 'leptons' emerged.

Dimensional frustration is a natural, almost inevitable, means of generating a 
hierarchy in QCD with five or more colors without resorting to contrived symmetry breaking
methods such as the Higgs field. Even a simplistic analysis such as this finds a rich
phenomenology, with further complexity expected at higher loop. 

The author thanks K.-I. Kondo for helpful discussions. This work was supported by a
fellowship from the Japan Society for the Promotion of Science (P05717), with hospitality
provided by the physics department of Chiba University.


\begin{thebibliography}{10}

\bibitem{S77}
G.K. Savvidy.
\newblock Infrared instability of the vacuum state of gauge theories and
  asymptotic freedom.
\newblock {\em Phys. Lett.}, B71:133, 1977.

\bibitem{F80}
H.~Flyvbjerg.
\newblock Improved qcd vacuum for gauge groups su(3) and su(4).
\newblock {\em Nucl. Phys.}, B176:379, 1980.

\bibitem{me07}
M.L. Walker.
\newblock Stability of the magnetic monopole condensate in three- and
  four-colour qcd.
\newblock {\em JHEP}, 01:056, 2007.

\bibitem{Cho80a}
Y.M. Cho.
\newblock A restricted gauge theory.
\newblock {\em Phys. Rev.}, D21:1080, 1980.

\bibitem{FN99c}
L.D. Faddeev and A.J. Niemi.
\newblock Partial duality in su(n) yang-mills theory.
\newblock {\em Phys. Lett.}, B449:214--218, 1999.

\bibitem{LZZ00}
Sheng Li, Yong Zhang, and Zhong-yuan Zhu.
\newblock Decomposition of su(n) connection and effective theory of su(n) qcd.
\newblock {\em Phys. Lett.}, B487:201--208, 2000.

\bibitem{NO78}
N.K. Nielsen and P.~Olesen.
\newblock An unstable yang-mills field mode.
\newblock {\em Nucl. Phys.}, B144:376, 1978.

\bibitem{H72}
J.~Honerkamp.
\newblock The question of invariant renormalizability of the massless
  yang-mills theory in a manifest covariant approach.
\newblock {\em Nucl. Phys.}, B48:269--287, 1972.

\bibitem{CmeP04}
Y.M. Cho, M.L. Walker, and D.G. Pak.
\newblock Monopole condensation and confinement of color in su(2) qcd.
\newblock {\em JHEP}, 05:073, 2004.

\bibitem{Cme04}
Y.M. Cho and M.L. Walker.
\newblock Stability of monopole condensation in su(2) qcd.
\newblock {\em Mod. Phys. Lett.}, A19:2707--2716, 2004.

\bibitem{CP02}
Y.M. Cho and D.G. Pak.
\newblock Monopole condensation in su(2) qcd.
\newblock {\em Phys. Rev.}, D65:074027, 2002.

\bibitem{K04}
K.-I. Kondo.
\newblock Magnetic condensation, abelian dominance and instability of savvidy
  vacuum.
\newblock {\em Phys. Lett.}, B600:287--296, 2004.

\bibitem{KKP05}
D.~Kay, A.~Kumar, and R.~Parthasarathy.
\newblock Savvidy vacuum in su(2) yang-mills theory.
\newblock {\em Mod. Phys. Lett.}, A20:1655--1662, 2005.

\bibitem{N74}
Y.~Nambu.
\newblock Strings, monopoles, and gauge fields.
\newblock {\em Phys. Rev.}, D10:4262, 1974.

\bibitem{M76}
S.~Mandelstam.
\newblock Vortices and quark confinement in nonabelian gauge theories.
\newblock {\em Phys. Rept.}, 23:245--249, 1976.

\bibitem{P77}
A.M. Polyakov.
\newblock Quark confinement and topology of gauge groups.
\newblock {\em Nucl. Phys.}, B120:429--458, 1977.

\bibitem{tH81}
G.~'t~Hooft.
\newblock Topology of the gauge condition and new confinement phases in
  nonabelian gauge theories.
\newblock {\em Nucl. Phys.}, B190:455, 1981.

\bibitem{KT00a}
K.~Kondo and Y.~Taira.
\newblock Non-abelian stokes theorem and quark confinement in su(3) yang-mills
  gauge theory.
\newblock {\em Mod. Phys. Lett.}, A15:367--377, 2000.

\bibitem{KT00b}
K.-I. Kondo and Y.~Taira.
\newblock Non-abelian stokes theorem and quark confinement in su(n) yang-mills
  gauge theory.
\newblock {\em Prog. Theor. Phys.}, 104:1189--1265, 2000.

\bibitem{Fbook87}
{P.H.~Frampton}.
\newblock {\em Quantum Field Theories}.
\newblock Benjamin-Cummings, California, 1987.

\bibitem{K06}
S.~Kato et~al.
\newblock Lattice construction of cho-faddeev-niemi decomposition and gauge
  invariant monopole.
\newblock {\em Phys. Lett.}, B632:326--332, 2006.

\bibitem{HF04}
Y.~Hatta and K.~Fukushima.
\newblock Linking the chiral and deconfinement phase transitions.
\newblock {\em Phys. Rev.}, D69:097502, 2004.

\bibitem{KLP01}
F.~Karsch, E.~Laermann, and A.~Peikert.
\newblock Quark mass and flavor dependence of the qcd phase transition.
\newblock {\em Nucl. Phys.}, B605:579--599, 2001.

\bibitem{YS82}
L.G. Yaffe and B.~Svetitsky.
\newblock First-order phase transition in the su(3) gauge theory at finite
  temperature.
\newblock {\em Phys. Rev. D}, 26(4):963--965, Aug 1982.

\bibitem{PW84}
R.D. Pisarski and F.~Wilczek.
\newblock Remarks on the chiral phase transition in chromodynamics.
\newblock {\em Phys. Rev.}, D29:338--341, 1984.

\end{thebibliography}

\end{document}